\documentclass[aps,prl,amsfonts,amssymb,epsfig,twocolumn,groupedaddress,showpacs]{revtex4} 
\usepackage{latexsym}
\usepackage[dvips]{graphicx}
\usepackage{epsfig} 
\usepackage{amsmath}

\providecommand{\be}{\begin{eqnarray}}
\providecommand{\enn}{\end{eqnarray}}

\begin{document}

\title{Characterizing electron entanglement in multiterminal mesoscopic conductors}

\author{Vittorio Giovannetti$^1$, Diego Frustaglia$^{1,2}$, Fabio Taddei$^1$, and
        Rosario Fazio$^{3,1}$}

\affiliation{
            $^1$NEST-CNR-INFM and Scuola Normale Superiore, I-56126 Pisa, Italy. \\
	    $^2$ Departamento de F\'isica Aplicada II, Universidad de Sevilla, E-41012 Sevilla, Spain.\\
            $^3$International School for Advanced Studies (SISSA), I-34014 Trieste, Italy}
\date{\today}
\begin{abstract}

We show that current correlations at the exit ports of a beam splitter 
can be used to detect electronic entanglement for a fairly general input state. 
This includes the situation where electron pairs can enter 
the beam splitter from the same port or be separated due to backscattering. 
The proposed scheme allows to discriminate between occupation-number and 
degree-of-freedom entanglement.

\end{abstract}


\maketitle

Generation, manipulation and detection of entangled electrons
is central for realizing integrated solid-state quantum computers. 
Among the several possibilities, a lot of attention has 
been devoted to the study of entanglement in multiterminal mesoscopic 
conductors  (see Refs.~\cite{beenakker,burkardR} for a review).
In this context, it was shown that 
entanglement between spatially separated electrons can be detected by 
means of a beam splitter (BS)~\cite{BLS00}. 
Indeed the BS, allowing the incoming (and possibly entangled) electrons 
to be interchanged, gives rise to two-particle interference effects. As a result, 
the symmetry of the incoming state influences the current-noise correlations 
at the exit ports. Bunching (enhanced) and anti-bunching (suppressed) behavior 
in the shot noise were predicted for spin singlet and triplet entangled states, 
respectively~\cite{BLS00}. The role of entanglement was later analyzed in the 
whole probability distribution of the current fluctuations (being
the noise power only its second moment)~\cite{taddei02}. Further analysis were 
subsequently performed in the presence of spin-orbit coupling~\cite{EBL02}, and 
for states generated in an Andreev double-dot entangler~\cite{SSB04}.
More recently, Burkard and Loss~\cite{burkardloss} found a bound for the 
entanglement of arbitrary mixed spin states through shot-noise 
measurements by applying a reduction mapping into Werner states~\cite{werner}.
We  generalized this result to multi-mode input states by introducing
an electronic Hong-Ou-Mandel interferometer \cite{gftf06}.

All the previous analysis rely on the assumption that only one 
electron per port
enters the analyzer. Most of the electronic 
entangler devices proposed so far~\cite{beenakker,burkardR}, however, generate 
states having a finite 
probability amplitude that two 
electrons enter the analyzer at the same input port~\cite{NOTAN}.
This gives rise to two distinct forms of entanglement: {\em occupation-number} 
and {\em electronic degree-of-freedom} entanglement \cite{SSR1,SSR2}.
Under this generalized initial condition the analysis of the entanglement 
is complicated by super-selection rules (SSR) induced by particle number 
conservation~\cite{beenakker,SSR1,SSR2,VC1}.

In this paper we present a detection strategy which addresses 
this more general situation. As in Ref.~\cite{gftf06}, it is based on the study of
current correlations at the exit ports of a BS as a function of  
controllable phase shifts. 
We shall show that, for the whole class of two-particle input states,
simple data processing of the measured current cross-correlators can be used to address separately
the various entanglement components.
Moreover, we also account for the case in which less than two electrons
can enter the interferometer due to backscattering.  
Our scheme of detection is particularly 
suitable when the entanglement is generated through superconducting-normal 
metal structures or multichannel non-interacting structures~\cite{BEN,RSL01}. 

\begin{figure}[t]
\begin{center}
\includegraphics[scale=0.24]{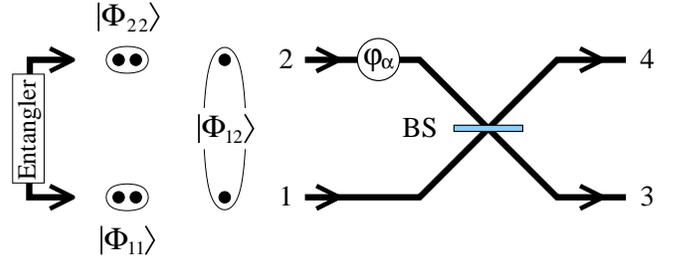}
\end{center}
\caption{Sketch of the analyzer. An external entangler prepares
electron pairs in the input state of Eq.~(\ref{statointot}) and 
injects them into ports $1$ and $2$. The electrons are allowed 
to enter the analyzer from the same port
(cases $|\Phi_{11}\rangle$ and $|\Phi_{22}\rangle$), 
from different ports (case $|\Phi_{12}\rangle$), 
or in any superposition of the previous cases.
Electrons propagating along lead
$2$ undergo an additional (orbital/spin-dependent) 
controllable phase shift (white circle in the figure)
before impinging on a $50\%$ beam splitter BS. 
Current correlations are measured at $3$ and~$4$.}
\label{skt}
\end{figure}

The proposed setup is described in Fig.~\ref{skt}. 
We start by considering 
a pair of electrons prepared by an external device 
(the ``entangler'' we want to monitor) at a given energy $E$ above the Fermi sea 
and injected into the interferometer at the input ports $1$ and $2$.
A generic input state has the form
\be
	|\Psi\rangle =
        \sin\theta \big( \cos\phi  \; |\Phi_{11}\rangle
        + \sin\phi \; |\Phi_{22}\rangle
        \big) + \cos\theta \; |\Phi_{12}\rangle
        \;, 
\label{statointot}
\enn
where $\theta,\varphi \in [0,\pi/2]$ and $|\Phi_{ij}\rangle$ describe two electrons 
of energy $E$ entering the BS from the $i$-th and $j$-th port respectively 
($i,j=1,2$). The latter  can be written as
	$|\Phi_{ij}\rangle= \sum_{\alpha,\beta} \Phi^{(ij)}_{\alpha,\beta} \, 
	a_{i,\alpha}^\dag(E) 
	a_{j,\beta}^\dag(E)|\O\rangle$,
with $a^\dag_{j,\alpha}(E)$ being the creation operator of an incoming 
electron in lead $j$ with orbital/spin degree-of-freedom index $\alpha$ 
and energy $E$, while $\Phi^{(12)}_{\alpha,\beta}$ and 
$\Phi^{(jj)}_{\alpha,\beta}=-\Phi^{(jj)}_{\beta,\alpha}$ 
satisfy the normalization condition
$\sum_{\alpha,\beta} |\Phi^{(12)}_{\alpha,\beta}|^2 =1$
and $\sum_{\alpha,\beta}|\Phi^{(jj)}_{\alpha,\beta}|^2 =1/2$.
In this notation $|\O\rangle \equiv|0\rangle_1\otimes |0\rangle_2$ 
is the vacuum representing the 
filled Fermi sea of the two leads.
The analysis of the entanglement contained in the state $|\Psi\rangle$,
can be done naturally by  bipartitioning the system with respect to 
the port labels $1$ and $2$.
For $\theta=0$, the state~(\ref{statointot}) reduces to the 
previously studied situation \cite{BLS00,burkardloss,gftf06}. 
There are however many
cases (e.g. the Andreev entangler~\cite{SSB04,RSL01})
where the incoming state instead has the form (\ref{statointot})
with $\theta\neq 0$. Our approach allows to analyze this situation as well.
Following Refs.~\cite{SSR1,SSR2}, the task is to identify both 
occupation-number and electronic degree-of-freedom entanglement. 
The first one coincides with the ``variance-EPR'' entanglement
of Ref.~\cite{SSR2} and with the ``fluffy bunny'' entanglement of
Ref.~\cite{SSR1}: It is present whenever we have a non trivial superposition 
among terms where each lead supports a different number of incoming electrons,
i.e. $|\Phi_{11}\rangle$, $|\Phi_{22}\rangle$, and $|\Phi_{12}\rangle$.
The second one, instead, coincides with the ``entanglement-EPR'' of Ref.~\cite{SSR2}
and originates from the component $|\Phi_{12}\rangle$ of Eq.~(\ref{statointot}): It 
is present when the leads $1$ and $2$ possess one electron each
and are entangled through the electronic orbital/spin modes $\alpha$.
Under the constraints imposed by the particle number SSR it can be shown~\cite{VC1,SSR2} 
that occupation-number and electronic degree-of-freedom entanglement are 
both quantum information resources which are crucial to solve specific tasks (examples 
are classical and quantum data hiding and standard quantum teleportation, respectively).

As shown in Fig.~\ref{skt}, electrons propagating along lead $2$ undergo an additional 
(channel/spin-dependent) controllable phase shift before impinging on a $50\%$ 
BS~\cite{notanew}.
For simplicity we assume that the BS is symmetric and does 
not suffer from backscattering. The BS is described by a scattering matrix 
$\hat{s}^{(\alpha)}$ defined through the relation 
$b_{j,\alpha}(E) 
= \sum_{j^\prime=1}^2 \; \hat{s}^{(\alpha)}_{j,j^\prime} 
\; a_{j^{\prime},\alpha}(E)
$, with $j=3,4$ and $b_{j,\alpha}(E)$ 
the annihilation operator of an outgoing 
electron in lead $j$ and channel $\alpha$ at energy $E$.
The scattering matrix is defined as
	$\hat{s}^{(\alpha)}
	\equiv \sqrt{1/2}\;\left(\begin{array}{cc}
	1&      
	e^{i \varphi_{\alpha}} \\
	1 &   - 
	 e^{i \varphi_{\alpha}} \\
	\end{array} \right) 
	$.
For definiteness we have imposed that no channel mixing occurs in the scattering process
at the BS. In the case of spin entanglement, this can be easily implemented by introducing
a BS conserving spin. For orbital entanglement, a spatial separation of the
orbital channels may be necessary.
Finally, the tunable phases $\varphi_{\alpha}$ are introduced 
at the input port 2 by some externally controlled parameters \cite{gftf06}.

The (dimensionless) current cross-correlator at the output ports $3$ and $4$ is
\begin{eqnarray}
	C_{34} 
\equiv \frac{h^2\nu^2}{2 e^2} 
\lim_{{\cal T} \to \infty} 
	\int_{0}^{\cal T} \!\!\! dt_1 
	dt_2 \; 
	\frac{\langle  I_{3}(t_1)  I_{4}(t_2) \rangle}{{\cal T}^2}
\;, \label{Xi} 
\end{eqnarray}
where~\cite{buttiker,lesovik},
\be
	I_j(t) \equiv \frac{e}{h\nu} \sum_{E,\omega, \alpha} e^{-i\omega t} 
 	[ b^\dag_{j,\alpha}(E)b_{j,\alpha}(E+\hbar\omega) \nonumber \\ 
	- a^\dag_{j,\alpha}(E)a_{j,\alpha}(E+\hbar\omega) ]
 \;
\label{current}
\enn
is the current operator of the $j$-th port ($j=3,4$). 
In these expressions the average $\langle \cdots \rangle$ is taken over the incoming 
electronic state, ${\cal T}$ is the measurement time, 
and $\nu$ is the density of states of the leads 
(a discrete spectrum has been considered to ensure a proper regularization of the 
current correlations).
$C_{34}$ is a measurable quantity which can be written in terms of the 
shot noise $S_{34}$ 
and the average current
$i_{3,4} = \int dt \langle I_{3,4}(t)\rangle/{\cal T}$ through the relation
$C_{34}= \frac{h\nu}{2 e^2} \left[ S_{34} + h\nu\; i_{3} i_{4} \right]$.
Differently from $S_{34}$, the
$C_{34}$ is linear in the input state of the system~\cite{NOTAQUI}.
Therefore, for any given density matrix 
$\rho =\sum_\ell p_\ell |\Psi_\ell\rangle\langle\Psi_\ell|$ 
with $|\Psi_\ell\rangle$ as in Eq.~(\ref{statointot}) we obtain
$C_{34}(\rho) =\sum_\ell p_\ell C_{34}(\Psi_\ell)$, where 
the correlators $C_{34}(\Psi_\ell)$ take the form
\begin{eqnarray}
	C_{34}(\Psi) = 
	\;[1+w \;\cos^2 \theta   
	\label{c2} 
	+ v\;  \sin^2\theta \sin(2 \phi)  ]/4 \;.
\end{eqnarray}
Here, $w$ and $v$ are real quantities satisfying $|w|,|v| \le 1$. 
They depend upon the interferometer phases $\varphi_\alpha$ 
and the input state parameters as
\begin{eqnarray}
w &=& \;  \sum_{\alpha,\beta} 
	\left(\Phi^{(12)}_{\alpha,\beta}\right)^* \; 
	\Phi^{(12)}_{\beta,\alpha} \; 
	e^{i ( \varphi_{\alpha} - \varphi_{\beta})} \nonumber \\
v &=&  2 \; \Re{e} \big[ \sum_{\alpha,\beta} 
	\left(\Phi^{(11)}_{\alpha,\beta} \right)^*
	\Phi^{(22)}_{\beta,\alpha} \;
	e^{i (\varphi_{\alpha} +\varphi_{\beta})}\big] \;.
\end{eqnarray}

Consider first the case where there is just one incoming electron per port, 
i.e., $\theta=0$ and $|\Psi\rangle = |\Phi_{12}\rangle$.
In this situation one has  $i_{3,4}=e/(h\nu)$ and Eq.~(\ref{c2}) gives the same Fano factor
$F_{34}=S_{34}/(2 e \sqrt{i_3 i_4})= [ w 
-1]/{4}$ of Ref.~\cite{gftf06},
where it was shown that
all separable mixtures of states
$|\Phi_{12}\rangle$ must exhibit  
non-negative values of $w$, i.e.  
non-negative  values of $C_{34}(\rho)-1/4$.

In the case where $\theta$ is generic, the same result can be obtained by 
relating the correlator $C_{34}$ to the entanglement of formation 
$E_f(\rho)$~\cite{bennett} of the input state.
This is done by applying generalized {\em twirling transformations} 
to map all density matrices for ports $1$ and $2$ into generalized
Werner states~\cite{TWIRL}.
The latter are constructed as in Ref.~\cite{gftf06} 
by introducing a joint orthonormal basis for ports $1$ and $2$
formed by the states $|\chi_k\rangle_1\otimes |\chi_k \rangle_2$ and
$|\Psi^{(\pm)}_{kk^\prime}\rangle= (|\chi_k\rangle_1\otimes 
|\chi_{k^\prime} \rangle_2 \pm |\chi_{k^\prime}
\rangle_1\otimes |\chi_k \rangle_2)/
\sqrt{2}$ with $k<k^\prime$. Here, the label $k$ enumerates all configurations 
with two or less particles in each port described, respectively, 
by vectors $|0\rangle_1$, 
$a_{1,\alpha}^\dag(E)|0\rangle_1$, 
$a_{1,\alpha}^\dag(E)a_{1,\beta}^\dag(E)|0\rangle_1$ 
and $|0\rangle_2$, 
$e^{-i\varphi_\alpha} a_{2,\alpha}^\dag(E)|0\rangle_2$,
$e^{-i (\varphi_\alpha +\varphi_\beta+\pi)} 
a_{2,\alpha}^\dag(E)a_{2,\beta}^\dag(E)|0\rangle_2$~\cite{NOTAW}.
This, indeed, generalizes the approach of 
Refs.~\cite{burkardloss,gftf06}, since we have introduced a 
larger basis that accounts for ports occupied by a different 
number of electrons.
Following the derivation of Refs.~\cite{burkardloss,gftf06},
it is easy to show that $E_f(\rho)$ can be lower  bounded 
by the quantity 
\begin{eqnarray}
W(\rho) = \sum_{kk^\prime} \langle \Psi_{k k^\prime}^{(-)} |
\rho |\Psi_{kk^\prime}^{(-)}\rangle /2 \label{DOPPIA}
\end{eqnarray}
through the inequality 
\be
E_f(\rho)\geqslant {\cal E}(W(\rho)) \label{inequ}\;,
\enn
with ${\cal E}(x) = H\left( \frac{1}{2}+ \sqrt{x(1-x)}\right)$
for $x\in [1/2,1]$ and null otherwise [here 
$H(x)=-x\log_2 x -(1-x)\log_2(1-x)$].
In our case, Eq.~(\ref{DOPPIA}) yields the identity
\begin{eqnarray}
W(\rho) = 1 - 2 C_{34}(\rho) 
\label{doppia}\;.
\end{eqnarray}
It follows that $C_{34}(\rho) <1/4$ 
implies $W(\rho)>1/2$ and hence, from Eq.~(\ref{inequ}), 
$E_f(\rho)>0$.
Therefore, we can conclude that also in the {\em general case} 
in which the two electrons can enter the same port, values 
of $C_{34}$ {\em smaller} than $1/4$ are  {\em direct evidence} of 
entanglement in the input state.

The sign of $C_{34} -1/4$, however, is not sufficient to  discriminate between 
occupation-number and electronic degree-of-freedom entanglement. 
In order to do so we use the symmetries
$w(\{\varphi_\alpha +\pi/2\}) =
w(\{\varphi_\alpha\})$ 
and 
$v(\{\varphi_\alpha +\pi/2\}) = - v(\{\varphi_\alpha\})$ 
to   define 
$C_{34}^{(\pm)}(\rho) 
\equiv [C_{34}(\rho ; \{\varphi_\alpha\}) \pm C_{34}(\rho ; 
\{ \varphi_\alpha +\pi/2\} ) ]/2 $.
Equation~(\ref{c2}) shows that $C_{34}^{(+)}$
depends only on the $|\Phi_{12}\rangle$
component of Eq.~(\ref{statointot}), while $C_{34}^{(-)}$
depends only upon the superposition among $|\Phi_{11}\rangle$
and $|\Phi_{22}\rangle$, i.e., 
\be
C_{34}^{(+)}(\Psi) 
&=& [1+w\; \cos^2\theta ]/4\;,
\label{cpiu} \\
C_{34}^{(-)}(\Psi) &=&  v\; 
	\sin^2\theta \sin (2\phi)  /4
\label{cmeno}\;.
\enn 
From the discussion of the $\theta=0$ case it follows that
the presence of electronic degree-of-freedom entanglement
can be detected by finding values of $\varphi_\alpha$ such that 
$C_{34}^{(+)}<1/4$~\cite{NOTAIMPORTANTE}. 
Vice-versa, we observe that
any  value of $C_{34}^{(-)}$ different from zero is indicative of occupation-number
entanglement in the system. Indeed, $C_{34}^{(-)} \neq 0$ is possible only 
for $\theta\neq 0$ and $\phi \neq 0,\pi/2$ which correspond to have a 
non-trivial superposition between terms with different local occupation number.
Moreover, one can relate $C_{34}^{(-)}$ to the superselection induced variance (SIV), 
$V(\Psi)=4\left( \langle \Psi| N_1^2|\Psi\rangle- \langle \Psi| N_1|\Psi\rangle^2 \right)$~\cite{SSR2}. This is a measure of occupation number entanglement,
where $N_1=\sum_{\alpha}a_{1,\alpha}^{\dagger} a_{1,\alpha}$ is the total number operator of particles in port 1.
It is easy to show that $V(\Psi)=(4 C_{34}^{(-)}/v)^2 + 4\sin^2{\theta}(\sin^2{\theta}\cos^2{\phi}+\cos^2{\theta})$, which allows to set a lower bound for the SIV, namely: $V(\Psi)\geq (4C_{34}^{(-)})^2$.

We finally discuss the case in which
the entangler suffers from backscattering,
a situation often encountered in proposed entanglers. In this case,
due to conservation of the total number of particles~\cite{beenakker,SSR1},
the most general input state in ports $1$ and $2$ is described by a
{\em convex combination} $R$ of density matrices 
$\rho^{\prime\prime}$, $\rho^\prime$ and $|\O\rangle\langle \O|$
having, respectively,  two, one or zero impinging electrons. It reads
$R = q^{\prime\prime} \; \rho^{\prime\prime} + q^\prime \;\rho^\prime+ (1-q^\prime-q^{\prime\prime}) \; |\O\rangle\langle\O|$,
where $q^{\prime\prime}$ is the probability that both electrons enter the ports 
(no backscattering occurred) and $q^{\prime}$ is the probability that only 
one of them enters the ports.
The entanglement of formation of the state~$R$ can be
bounded as before. In this case, however, Eqs.~(\ref{DOPPIA}) and (\ref{doppia})
give
\be
W(R)&=& q^{\prime\prime} W(\rho^{\prime\prime})+ q^{\prime} W(\rho^{\prime}) 
\geqslant  q^{\prime\prime} W(\rho^{\prime\prime}) \nonumber \\
&=&  q^{\prime\prime} 
\big[1 - 2 C_{34}(\rho^{\prime\prime}) \big] =\label{doppiowdir}
q^{\prime\prime} - 2 C_{34}(R)\;,
\enn
where we used the fact that $W(|\O\rangle\langle\O|)=0$ and the linearity
of $C_{34}$, yielding
$C_{34}(R) = q^{\prime\prime} \; C_{34} (\rho^{\prime\prime})$ (notice that since $\rho^\prime$ 
and $|\O\rangle$ have less than two electrons they do not  contribute
 to the current cross-correlator).
From  the monotonicity  of the function ${\cal E}(x)$ we finally obtain
\begin{eqnarray}
E_f(R) \geqslant {\cal E}(q^{\prime\prime} -2 C_{34}(R)) \;. \label{efdir}
\enn
Once the transmission probability $q^{\prime\prime}$ of the setup is known,
Eq.~(\ref{efdir}) provides an estimate of the entanglement of formation of the system.
Notice that, differently from the case discussed in the first part of the paper, 
the state~$R$ is a mixture of terms with different total number of particles.
In this case the standard entanglement of formation $E_f(R)$ 
measures the amount of entanglement needed to create the state $R$ without
taking into account the constraints posed by particle-number conservation
(i.e., considering possible decompositions of $R$ including coherent superpositions of 
states with different particle number).
For this reason Schuch, Verstraete and Cirac~\cite{SSR2} have proposed to replace the 
quantity $E_f(R)$ with the {\em SSR-entanglement of formation}. This is defined as 
$E_f^{(SSR)}(R) = \min_{p_i,\psi_i} \sum_i p_i E_f(\psi_i)$, 
where the minimization is performed over all decompostion of $R$ with $|\psi_i\rangle$ 
being eigenstates of the particle-number operator.
By construction, $E_f^{(SSR)}(R)$ is always greater or equal to $E_f(R)$. 
In our case it reads
\be
E_f^{(SSR)}(R) =
 q^{\prime\prime} \; E_f(\rho^{\prime\prime}) + q^{\prime} \; E_f(\rho^{\prime}) \geqslant 
 q^{\prime\prime} \; E_f(\rho^{\prime\prime}) \label{dise} \;,
\enn
where we used $E_f(|\O\rangle\langle\O|)=0$. The rhs term of this expression can now
be lower bounded by noticing that 
$\rho^{\prime\prime}$ represents a density matrix formed by vectors of the 
form~(\ref{statointot}). Employing Eqs.~(\ref{inequ}) and (\ref{doppia}) we find
\be
E_f^{(SSR)}(R) 
\geqslant q^{\prime\prime} \; {\cal E}\big(1-2C_{34}(R)/q^{\prime\prime}\big)\;, \label{dise1}
\enn
where we used, again, the relation
$C_{34}(R) = q^{\prime\prime} \; C_{34} (\rho^{\prime\prime})$
by linearity of $C_{34}$.
As in the case of Eq.~(\ref{efdir}), 
once the probability $q^{\prime\prime}$ is known,
Eq.~(\ref{dise1}) provides an estimate of the SSR-entanglement of formation of the system.

In the remaining part of the paper we make a connection 
between the results obtained so far and some 
proposed electronic entanglers. 
The situation we consider applies naturally to
the case of superconductor-based entanglers. It has 
been shown~\cite{SSB04} that Cooper pairs can be injected into normal metal
ports 1 and 2 through a (double) tunnel barrier in the form of a wave packet 
$|\Psi \rangle = \sum_E |\Psi_E\rangle$. Here, the states $|\Psi_E\rangle$ 
have the form given in  Eq.~(\ref{statointot}), with the difference that the two 
electrons do not possess the same energy.
In fact, they have opposite sign with respect to the 
chemical potential of the superconductor, instead, such that 
$|\Phi_{ij} \rangle=\sum_{\alpha,\beta} \Phi^{(ij)}_{\alpha,\beta}\, a_{i,\alpha}^\dag(E) a_{j,\beta}^\dag(-E)|\O\rangle$. 
When calculating the cross-correlator $C_{34}$ of Eq.~(\ref{Xi}) one notices that, 
actually, the energy labels $E$ and $-E$ can be effectively incorporated into the mode 
labels $\alpha$ and $\beta$ in $|\Phi_{ij} \rangle$. This is possible thanks to the 
limit $\cal{T}\rightarrow \infty$ in Eq.~(\ref{Xi}), which allows only zero-frequency contributions 
($\omega=0$) to the current operators (\ref{current}). 
Two-particle wave packets produced by superconductor-based entanglers can be
then described by Eq.~(\ref{statointot}) [up to a normalization factor] once the 
energy index is accounted for by the electronic degree-of-freedom one. Moreover, notice 
that introducing phases $\varphi_\alpha$ that can discriminate between different 
energies would eventually lead to the characterization of energy entanglement in 
the injected Cooper pairs.

Regarding non-interacting electron entanglers, at 
zero temperature they typically produce states of the form 
$|\Psi \rangle = \prod_E |\Psi_E\rangle$ with $|\Psi_E\rangle$ as 
in Eq.~(\ref{statointot}). The energy product corresponds to a discretization with 
interval $\delta E=h/\cal{T}$ in an energy window $eV$, where $V$ is a voltage 
difference applied between two electron reservoirs~\cite{beenakker}. 
The state $|\Psi \rangle$ 
describes $N=eV/\delta E$ incoming electron pairs. 
In the case $N>1$, the cross-correlator (\ref{Xi}) exhibits extra terms 
scaling with $N(N-1)$ 
which do not allow to separate the occupation-number entanglement from the 
electronic degree-of-freedom one -- details shall be given elsewhere.
However, the problem 
may be solved by introducing small measurement times instead, 
such that $eV \ll \delta E$  and $N \rightarrow 1$. 
It can be shown that a 
calculation of the the cross-correlator (\ref{Xi}) in these conditions leads
to equivalent results.

This work was supported by the European Commission (grants RTN Spintronics, RTNNANO 
and EUROSQIP), by MIUR-PRIN, by the TRIL-ICTP program, by the "Ram\'on y Cajal" program of the Spanish Ministry of Education and Science, 
and by the Quantum Information 
research program of the "Ennio De Giorgi" Mathematical Research Center of SNS.



\begin{thebibliography}{99}
\bibitem{notanew}
Our method requires to know the transparency of the beam splitter. The choice of a 50\% transparency is not fundamental, but it allows us to simplify the expressions. It corresponds to the ideal situation in which the entanglement detection capability is maximized.
\bibitem{NOTAQUI}
This is because
for $\theta \neq 0$ in Eq.~(\ref{statointot}), 
the average currents $i_{3,4}$ are non trivial functions of the phases 
$\varphi_\alpha$ and of the input state $|\Psi\rangle$. Instead,
for $\theta=0$ the $i_{3,4}$ are
constant quantities which do not depend on 
$|\Psi\rangle = |\Phi_{12}\rangle$. This is the only case where 
also $S_{34}$ 
is  linear (see Ref.~\cite{gftf06}).
\bibitem{NOTAW}Ambiguity in the definition of the joint
vectors $|\chi_k\rangle_1\otimes |\chi_{k^{\prime}}\rangle_2$ 
can be avoided by
simply establishing a convention for the ordering of 
the creators operators $a_{j,\alpha}^\dag$.
\end{thebibliography}
\end{document}